\documentstyle[12pt]{article}
\newcommand{\nt}{\mbox{$\bar{n}(t)$}}
\newcommand{\Pti}{\mbox{$P_{\infty}$}}
\newcommand{\gti}{\rightarrow  \infty}
\newcommand{\Rt}{\mbox{$\bar{R}^{2}(t)$}}
\newcommand{\ave}[1]{\mbox{$\langle #1 \rangle$}}
\newcommand{\nuv}{\nu_{\parallel}} 
\newcommand{\nuh}{\nu_{\perp}}
\newcommand{\rhoss}{\mbox{$\bar{\rho}$}} 

\def\figlsfar{Fig.~1}
\def\figlsclose{Fig.~2}
\def\figrhovsl{Fig.~3}
\def\figchivsl{Fig.~4}
\def\figtauvsl{Fig.~5}
\def\figrhoinit{Fig.~6}
\def\figsupscale{Fig.~7}
\def\figpinf{Fig.~8}
\def\figdsinf{Fig.~9}
\def\figninf{Fig.~10}
\def\fignpsub{Fig.~11}
\def\figr2sub{Fig.~12}
\def\figrhoss{Fig.~13}
\def\figchiss{Fig.~14}
\def\figrhovsh{Fig.~15}
\def\figlsnabs{Fig.~16}

\def\tableexp{Table~I}

\title{Critical Exponents for Branching Annihilating Random Walks 
with an Even Number of Offspring}

\author{Iwan Jensen \\  Department of Mathematics, 
The University of Melbourne, \\ Parkville, Victoria 3052,  
Australia \\ e-mail: iwan@maths.mu.oz.au}

\maketitle

\begin{document}
\setcounter{page}{0}
\thispagestyle{empty}

\begin{abstract} 
Recently, Takayasu and Tretyakov [Phys. Rev. Lett. {\bf 68}, 
3060 (1992)], studied branching annihilating random walks (BAW)
with $n=1$-5 offspring. These models exhibit  a continuous
phase transition to an absorbing state. For odd $n$ the
models belong to the universality class of directed percolation.
For even $n$ the particle number is conserved modulo 2 and
the critical behavior is not
compatible with directed percolation. In this article I study
the BAW with $n=4$  using time-dependent simulations and
finite-size scaling obtaining precise estimates for various critical
exponents. The results are consistent with the conjecture:
$\beta/\nuh = \frac{1}{2}$, $\nuv/\nuh = \frac{7}{4}$, $\gamma = 0$,  
$\delta = \frac{2}{7}$, $\eta = 0$, $z = \frac{8}{7}$, and
$\delta_{h} = \frac{9}{2}$.
\end{abstract}


\newpage

\section{Introduction \label{sec-intro}}

Extensive studies of various nonequilibrium models exhibiting a 
second-order phase transition to a {\em unique} absorbing state have
revealed that directed percolation (DP) \cite{dpbook}-\cite{pgdp} is 
the generic critical behavior of such models. Other well-known models 
belonging to the DP universality class are the contact 
process \cite{harris}-\cite{durrett}, Schl\"{o}gl's first and second 
models \cite{schlogl}-\cite{pg2}, and Reggeon field theory 
(RFT) \cite{torre,brower}. Numerous models studied in recent years 
demonstrate the robustness of this universality class against a wide
range of changes in the local kinetic rules such as multi-particle
processes \cite{bbc}--\cite{tripcrea}, diffusion \cite{rondif} and 
changes in the number of components \cite{zgb}. Very recently it
was shown that even models with infinitely many absorbing states 
may exhibit DP critical behavior \cite{pcpart}.

Recently, Takayasu and Tretyakov \cite{bawprl}, studied the 
branching annihilating random walk (BAW) with $n=1$-5 offspring.
The BAW is a lattice model in which each site is either empty or
occupied by a single particle. The evolution rules are quite simple
and each 'elementary' step starts by chosing a particle at random.
With probability $p$ the particle jumps to a randomly chosen nearest 
neighbor. If this site is already occupied the particles annihilate 
mutually. With probability $1-p$ the particle produces $n$ offspring, 
which are placed on the closest neighboring sites. When an offspring 
is created on a site already occupied both particles annihilate
leaving behind an empty site. In one dimension for $n=1$ it has been 
shown \cite{bawbramson} that the BAW has an active steady-state for 
sufficiently small $p$. Computer simulations revealed that the phase 
transition from the active state to the absorbing state is continuous 
\cite{bawprl}. BAW's with an odd number of offspring include a 
combination of diffusion and various multi-particle processes. One 
would therefore expect, bearing in mind the robustness of DP critical 
behavior, that the transition should belong to the universality 
class of directed percolation. The steady-state concentration of 
particles $\rhoss$ (which is the appropriate order parameter) 
decays as:

\begin{equation}
 \rhoss \propto |p_{c} - p|^{\beta}, \label{eq:rhoss}
\end{equation}

in the supercritical regime ($p < p_{c}$), where $\beta$ is the order 
parameter critical exponent. Estimates for $\beta$, obtained from 
computer simulations, were however only marginally consistent with 
directed percolation. Takayasu and Tretyakov \cite{bawprl} 
found $p_{c} = 0.108 \pm 0.001$ and $\beta = 0.32 \pm 0.01$ 
which should be compared to the value $\beta = 0.2769 \pm 
0.0002$ \cite{tdpjsp} for the one-dimensional contact process. 
For $n = 3$ and $n = 5$ they found that $p_{c} = 0.461 \pm 0.002$ 
and $0.718 \pm 0.001$, respectively, with $\beta = 0.33 \pm 0.01$ 
in both cases. Time-dependent computer simulations \cite{iwanbaw} 
for $n=1$ and 3 yielded estimates for three critical exponents in 
good agreement with directed percolation, thus supporting the notion 
that BAW's with an odd number of offspring belong to the DP 
universality class. This conclusion has been confirmed recently by 
a study \cite{inuicam} of the BAW with $n=1$ yielding a value of 
$\beta$ consistent with DP, using mean field cluster expansions and 
the coherent anomaly method \cite{cam}.

When $n$ is {\em even} the number of particles is conserved modulo 2 
and the absorbing state can only be reached if one initially has 
an even number of particles. For $n = 2$ the model does not 
have an active steady state \cite{bawsudbury,bawtakayasu} whereas for 
$n=4$ it was found \cite{bawprl} that  $p_{c}=0.72\pm 0.01$ and 
$\beta = 0.7\pm 0.1$, which clearly places the model outside the DP 
universality class. Grassberger {\em et al.} \cite{pgkink} studied a 
model, involving the processes $X \rightarrow 3X$ and 
$2X \rightarrow 0$, from which it is obvious that once again the 
number of particles is conserved modulo 2. Steady-state and 
time-dependent computer simulations yielded non-DP values for various 
critical exponents (this model and the results will be described in 
more detail in Section~\ref{sec-kink}). The conservation of 
particle number modulo 2 
is probably responsible for the non-DP behavior. Support for 
this idea was provided by a recent study \cite{bawann} of a modified 
version of the BAW with $n=4$. In this study spontaneous annihilation 
of particles was included, thus destroying the conservation of 
particles modulo 2. Results from computer simulations clearly showed 
that adding spontaneous annihilation changes the critical behavior 
from non-DP to DP, even for very low rates of spontaneous annihilation
\cite{bawann}.

As non-DP behavior is so rare it is clearly of great interest to study
BAW's with an even number of offspring in greater detail. In this 
paper I present results from extensive computer simulations of the 
one-dimensional BAW with $n=4$ using time-dependent simulations and 
finite-size scaling analysis of steady-state simulations.

\section{Time-dependent simulations \label{sec-td}}

Earlier studies \cite{pgdp,torre,tripann,tripcrea,cp3d} have 
revealed that time-dependent simulations is a very efficient method 
for determining critical points and exponents for models with a 
continuous transition to an absorbing state. The general idea of 
time-dependent simulations is to start from a configuration which 
is very close to the absorbing state, and then follow the 'average' 
time evolution of this configuration by simulating a large ensemble 
of independent realisations. In the simulations presented here I 
always started, at $t=0$, with 2 occupied nearest neighbor sites 
placed at the origin. I then made a number, $N_{S}$, of independent 
runs, typically $5 \times 10^{5}$, for different values of $p$ in 
the vicinity of $p_{c}$. As the number of particles is very small 
an efficient algorithm may be devised by keeping a list of occupied 
sites. In each elementary step a particle is draw at random from this 
list and the processes are performed according to the rules given 
earlier. Before each elementary change the time variable is 
incremented by $1/n(t)$, where $n(t)$ is the number of particles on 
the lattices prior to the change. This makes one timestep equal to 
(on the average) one attempted update per lattice site. Each run had 
a maximal duration, $t_{M}$, of 5000  time steps. I measured the 
survival probability $P(t)$ (the probability that the system had not 
entered the absorbing state at time $t$), the average number of 
occupied sites $\bar{n}(t)$, and the average mean square distance of 
spreading $\bar{R}^{2}(t)$ from the center of the lattice. Notice 
that $\bar{n}(t)$ is averaged over all runs whereas $\bar{R}^{2}(t)$ 
is averaged only over the surviving runs. From the scaling ansatz for 
the contact process and similar models \cite{pgdp,torre,cp3d} it 
follows that the quantities defined above are governed by power laws 
at $p_{c}$ as $t \gti$ 
 
\begin{eqnarray}
P(t) & \propto & t^{-\delta }\Phi ( \Delta t^{1/\nuv}), 
\label{eq:sp} \\
\bar{n}(t) & \propto & t^{\eta }\Psi ( \Delta t^{1/\nuv}) , 
\label{eq:nt} \\
\bar{R}^{2}(t) & \propto & t^{z}\Theta ( \Delta t^{1/\nuv}), 
\label{eq:Rt}
\end{eqnarray}
 
where $\Delta = |p_{c} - p|$ is the distance from the critical point 
and $\nuv$ is the correlation length exponent in the time direction. 

If the scaling functions $\Phi$, $\Psi$, and $\Theta$ are 
non-singular at the origin it follows that $P(t)$, \nt, and \Rt\ 
behave as pure power-laws at $p_{c}$ for $t \gti$. In log-log plots
of $P(t)$, $\bar{n}(t)$ and $\bar{R}^{2}(t)$ as a function of $t$ 
one should see (asymptotically) a straight line at $p=p_{c}$. The 
curves will show positive (negative) curvature when $p<p_{c}$
$(p>p_{c})$. This enables one to obtain accurate estimates for $p_{c}$. 
The asymptotic slope of the (critical) curves define the dynamic 
critical exponents $\delta,\ \eta$ and $z$. Generally one has to 
expect corrections to the pure power law behavior so that $P(t)$ 
is more accurately given as \cite{pgdp}
 
\begin{equation}
P(t) \:\propto \:t^{-\delta }(1 \ + \ at^{-1} \ 
+ \ bt^{-\delta '}\ + \ \cdots \ )
\end{equation}
 
and similarly for $\bar{n}(t)$ and $\bar{R}^{2}(t)$. More 
precise estimates for the critical exponents can be obtained 
by looking at local slopes
 
\begin{equation}
  -\delta (t) \:=\:\frac{\ln[P(t)/P(t/m)]}{\ln(m)}, \label{eq-ls}
\end{equation}
 
and similarly for $\eta (t)$ and $z(t)$; in this work I used $m=5$. 
The local slope $\delta (t)$ behaves as \cite{pgdp}
 
\begin{equation}
  \delta(t)\:=\:\delta \ + \ at^{-1} \ + \ b\delta 't^{-\delta '}\ 
     + \ \cdots
\end{equation} 
 
and similar expressions for $\eta (t)$ and $z(t)$. Thus in a plot of 
the local slopes {\em vs} $1/t$ the critical exponents are given by 
the intercept of the curves for $p_{c}$ with the $y$-axis. The 
off-critical curves often have a very notable curvature, i.e., one 
will see the curves for $p > p_{c}$ veering downward while the curves 
for $p < p_{c}$ veer upward.

The local slopes thus provides us with a very efficient means for 
estimating both the location of the critical point and the values of
various critical exponents. In \figlsfar\ I have plotted the local 
slopes for various values of $p$. The maximal number of timesteps
$t_{M}=5000$ and the number of samples $N_{S}=5\times10^{5}$ except
for $p=0.721$ where $N_{s}=2\times10^{6}$. One clearly sees that 
the curves for $p=0.715$ veer upwards and the curves for $p=0.727$
veer downward and these values of $p$ are thus off-critical. For 
$p=0.724$ the curve for $-\delta(t)$ veers downward at the end as 
$1/t \rightarrow 0$. I therefore conclude that $p_{c}<0.724$. The 
curves for $p=0.718$ does not exhibit pronounced curvature and is thus 
close to $p_{c}$, however $\eta(t)$ does seem to veer upwards so 
$p_{c}$ is probably larger than 0.718. All in all it seems that 
$p=0.721$ is the value closest to $p_{c}$ leading to the estimate 
$p_{c}=0.721 \pm 0.003$, which is in excellent agreement with 
the estimate $p_{c}=0.72 \pm 0.01$ obtained by Takayasu and 
Tretyakov \cite{bawprl}. In order to obtain a better estimate for 
$p_{c}$ and the critical exponents I performed extensive 
simulations for  $p = 0.7210$, 0.7215  and 7220, with $t_{M}=5000$ 
and $N_{S}=2\times 10^{6}$. The resulting local slopes are shown 
in \figlsclose. None of the curves  
exhibit pronounced curvature, however the estimate $\eta = 0$ 
obtained for $p = 0.7215$ is to preferred (if only for aesthetical 
reasons) over an estimate that is very close to 
zero. All in all I conclude that $p_{c} = 0.7215 \pm 0.0005$. 
From the intercept of the critical curves with the $y$-axis I obtain 
the estimates $\delta = 0.286 \pm 0.001$, $\eta = 0 \pm 0.001$, 
and $z = 1.143 \pm 0.003$. These result certainly suggests that 
$\eta = 0$ is an {\em exact} result. One might thus hope that the 
other exponets are given by simple fractions, and indeed I find that 
$\delta=\frac{2}{7}=0.28571\ldots$ and $z=\frac{8}{7}=1.14285\ldots$ 
are consitent with the simulation results. Note that both the 
conjectured exact values and the estimates obey the hyper-scaling 
relation \cite{pgdp,torre,cp3d} 

\begin{equation}
	4\delta + 2\eta = dz, \label{eq:hsr}
\end{equation}

which lends further support to the validity of the asumptions made
above.

\section{Finite-size scaling analysis \label{sec-fss}}

The concepts of finite-size scaling \cite{fss}, 
though originally developed for equilibrium systems, are 
also applicable to nonequilibrium second-order phase transitions. 
Aukrust {\em et al.} \cite{aukrust} showed how finite-size scaling 
can be used very successfully to study the critical behavior of 
nonequilibrium systems exhibiting a continuous phase transition to 
an absorbing state. As in equilibrium second-order phase transitions 
one assumes that the (infinite-size) nonequilibrium system features 
length- and time-scales which diverge at criticality as 

\begin{equation}
   \xi(p) \propto |p_{c} - p|^{-\nuh}, \label{eq:xi} 
\end{equation}
and 
\begin{equation}
  \tau (p) \propto |p_{c} - p|^{-\nuv}, \label{eq:tauss}
\end{equation}

where $\nuh$ ($\nuv$) is the correlation length exponent in the space 
(time) direction. 

\subsection{Static behavior \label{sec-fssstatic}}

One expects finite-size effects to become important when the 
correlation length $\xi(p) \sim L$. The basic finite-size scaling 
ansatz is that various physical quantities depend on system-size only 
through the scaled length $L/\xi(p)$, or equivalently through the 
variable \mbox{$(p_{c}-p)L^{1/\nuh}$}. Thus we assume that the order 
parameter depends on system size and distance from the critical 
point as:
	     
\begin{equation}
  \rho_{s}(p,L) \propto L^{-\beta/\nuh} 
  f((p_{c}-p)L^{1/\nuh}), \label{eq:rhofss}
\end{equation}
such that at $p_{c}$
 \begin{equation}
  \rho_{s}(p_{c},L) \propto L^{-\beta/\nuh}, \label{eq:rhocr}
\end{equation}

and $f(x) \propto x^{\beta} \ \ \mbox{for} \ \ x \gti$,
so that Eq.~(\ref{eq:rhoss}) is recovered when $L \gti$ in the 
critical region. In $\rho_{s}$, and other quantities, the subscript 
$s$ indicates an average taken over the {\em surviving} samples, 
i.e., the average includes only those samples which have not yet 
entered the absorbing state. The restriction to surviving samples 
ensures that the system enters a 'quasisteady state' in which
a quantity such as $\rho_{s}$ becomes constant after a 
relative short transient time \cite{aukrust,pcpart}. In the 
supercritical regime $\rho_{s}$ should approach a nonzero asymptote 
for $L \gg \xi(p)$, while in the subcritical regime one expects 
$\rho_{s}$ to decay faster than a power-law. Thus $p_{c}$ may be 
determined as the value yielding a straight line in log-log plot of 
$\rho_{s}$ versus $L$. However, since the time-dependent simulations
already yielded an accurate estimate of $p_{c}$ I have only performed
the steady-state simulations for $p=p_{c}=0.7215$ with various
values of the system-size $L$. \figrhovsl\ shows a log-log plot of 
the average concentration of particles $\rho_{s}(p_{c},L)$ in the 
quasisteady state as a function of $L$. The number of timesteps 
$t_{M}$ and independent samples $N_{S}$ varied from $t_{M}=200$, 
$N_{S}=200,000$ for $L=16$ to $t_{M} = 5,000,000$, $N_{S}=100$ for 
$L=16,384$. As can be seen the data falls very nicely on a straight
line, form the slope of this line I obtain the estimate 
$\beta/\nuh = 0.50 \pm 0.01$. This result clearly suggests that
$\beta/\nuh = \frac{1}{2}$.

Another exponent estimate can be obtained from the fluctuations, 

\begin{equation}
  \bar{\chi} = L^{d}(\ave{\rho^{2}} - \ave{\rho}^{2}) \propto
	 |p_{c} - p|^{-\gamma}, \label{eq:chiss}
\end{equation}

where $L$ is the linear extension of the system. Eq.~(\ref{eq:chiss})
thus leads to the following finite-size scaling ansatz:

\begin{equation}
 \chi_{s}(p,L) \propto L^{\gamma/\nuh} 
    g((p_{c} - p)L^{1/\nuh}), \label{eq:chifss}
\end{equation}
and
\begin{equation}
  \chi_{s}(p_{c},L) \propto L^{\gamma/\nuh}. \label{eq:chicr}
\end{equation}

When plotting the data for $\chi$ on a log-log scale the curve had
no pronounced slope showing that $\gamma \simeq 0$ suggesting that 
the fluctuations grows only logarithmically with system size.
\figchivsl\ shows a log-lin plot of the susceptibility 
$\chi_{s}(p_{c},L)$ as a function of $L$. The data falls more or 
less on a straight line (with large error-bars for large 
$L$-values) showing that $\gamma$ indeed is zero. Note that this 
result through the scaling relation \cite{aukrust,pcpart}

\begin{equation}
  \gamma = d\nuh - 2\beta, \label{eq:gammascaling}
\end{equation}

leads to the {\em exact} result $\beta/\nuh = \frac{1}{2}$ in full
agreement with the simulation results for $\rho_{s}$.

\subsection{Dynamical behavior \label{sec-fssdyn}}

Additional exponents may be obtained from the dynamical behavior
of the system. In this study I define a characteristic time,
$\tau(p,L)$, as the time it takes for half the sample to 
enter the absorbing state. From Eq.~(\ref{eq:tauss}) we obtain the 
following finite-size scaling form:

\begin{equation}
  \tau(p,L) \propto L^{y}h((p_{c} - p)L^{1/\nuh}),
  \label{eq:taufss}
\end{equation}

where $y = \nuv/\nuh$. At $p_{c}$ we thus have

\begin{equation}
  \tau(p_{c},L) \propto L^{y}. \label{eq:taucr}
\end{equation}

In \figtauvsl\ I have plotted, on a log-log scale,  
$\tau_{s}(p_{c},L)$ as a function of $L$. Again we see a nice
straight line this time with slope $\nuv/\nuh = 1.75\pm 0.01$,
which leads to the conjecture $\nuv/\nuh = \frac{7}{4}$. This
value agrees with the scaling relation \cite{pgdp},

\begin{equation}
 z = 2\nuh/\nuv. \label{eq:zscaling}
\end{equation}

One may also study the dynamical behavior by looking at the time 
dependence of $\rho_{s}(p_{c},L,t)$. For $t \gg 1$ and $L \gg 1$ 
one can assume a scaling form

\begin{equation}
  \rho_{s}(p_{c},L,t) \propto L^{-\beta/\nuh}f(t/L^{y}). 
  \label{eq:rhofssall}
\end{equation}

At $p_{c}$ the system shows power-law behavior for $t < L^{y}$ 
before finite-size effects become important. Thus for $L \gg 1$ and 
$t < L^{y}$, $\rho(p_{c},L,t) \propto t^{-\theta}$. From 
Eq.~(\ref{eq:rhofssall}) it then follows that 

\begin{equation}
   \theta = \beta/(\nuh y) = \beta/\nuv = \delta. \label{eq:thetasca}
\end{equation}

\figrhoinit\ shows the short-time evolution of the concentration of 
particles at $p_{c}$. The slope of the line drawn in the figure is
$\theta=\frac{2}{7}$ and this value seems to be in excellent accord 
with the simulation data.

\section{Determining $\beta$, $\nuv$, and $\nuh$}

The numerical values of $\nuv$ and $\nuh$ turns out to be quite large.
It is therefore difficult to obtain accurate estimates for these 
exponents and $\beta$ since the correlation-time and -length grows 
rapidly as one approaches the critical point. The exponent estimates 
presented in this section are thus marked by far greater uncertainties
than those of the preciding sections.  

\subsection{Time-dependent simulations}

Estimates for $\beta$ and $\nuv$ can be obtained from time-dependent 
simulations in the super- and subcritical regions \cite{cp3d}. In 
the {\em supercritical} region the number of indepent runs $N_{S}$ 
and the maximal duration $t_{M}$ of each run varied from 
$N_{S}=15,000, \ t_{M}=1,000,000$ closest to $p_{c}$, to 
$N_{S} = 100,000,\ t_{M} = 100$ furthest away from $p_{c}$. One
may analyze the data by utilizing the  scaling behavior expressed 
in Eqs.~(\ref{eq:sp})-(\ref{eq:Rt}). First we rewrite the equations 
by changing the variable to $x=\Delta t^{1/\nuv}$. The expression 
for the survival probability then becomes (after noting that
$t^{-\delta} = \Delta^{\nuv \delta} x^{-\nuv \delta}$):

\begin{equation}
  P(x) \ \propto \ \Delta^{\nuv \delta} x^{-\nuv \delta} \Phi (x).
\end{equation}

The expressions for \nt\ and \Rt\ can be rewritten in a similar
manner. In plots of $\Delta^{- \nuv \delta} P(t)$,
$\Delta^{\nuv \eta} \nt$ or $\Delta^{\nuv z} \Rt$ {\em vs} $x$
we should see the data collapse onto single curves. Since
I am fairly certain that the estimates for $p_{c}$, $\delta$, $\eta$,
and $z$ are very accurate there is really only one free variable
$\nuv$ to be determined in the scaling analysis. \figsupscale\ 
shows the results with $p_{c}=0.7215$, $\delta=\frac{2}{7}$, 
$\eta=0$, and $z = \frac{8}{7}$, while the 'free' variable 
$\nuv=3.25$, which leads to a decent data collapse. The curves that 
does not collapse onto the common curve are those furthest away from 
$p_{c}$ and thus those for which corrections to scaling are expected
to be the strongest. The value $\nuh = 3$ also leads to a decent 
collapse, but now the deviating curves are those closest to $p_{c}$ 
which should rule out this value. $\nuv = 3.5$ is also a possibility,
but the scaling collapse is less impressive in this case. Due to the
fact that the simulations could not be extended closer to $p_{c}$
(the demands in CPU time simply become to great) it is not possible
to rule this value out entirely. I thus conclude that 
$\nuv=3.25\pm 0.25$ though the lower end of this range is highly
unlikely. Using the finite-size results $\nuv/\nuh=\frac{7}{4}$ and
$\beta/\nuh=\frac{1}{2}$ we see that $\nuv=\frac{13}{4}$ gives the
estimates $\nuh = \frac{13}{7}$ and $\beta=\frac{13}{14}$, whereas
$\nuv = \frac{7}{2}$ yields $\nuh = 2$ and $\beta = 1$. The latter set 
of values is certainly aesthetically more appealing, but as I will 
argue below the simulations does seem to favor the former values.

$\beta$ and $\nuv$ can also be estimated directly from the 
asymptotic behavior of $P(t)$ and \nt. We see that by setting 
$\phi(y) \ = \ y^{-\delta\nuv}\Phi(y)$ we may rewrite 
Eq.~(\ref{eq:sp}) as

\begin{equation}
  P(t) \ \propto \ \Delta^{\nu \delta}\phi(\Delta t^{1/\nuv}).
   \label{eq:pts}
\end{equation}

Since the system is in the supercritical region there must be a
finite chance of survival; were this not the case any configuration
would eventually die out, contrary to our knowledge that the
system has an active steady state in this regime. Thus since
$\Pti \equiv \lim_{t \gti}P(t)$ is finite, $\lim_{y \gti} \psi(y)$
is finite too, and we get:

\begin{equation}
\Pti \ \propto \ \Delta^{\nuv \delta}. \label{eq:usp}
\end{equation}

For the contact process it has been shown \cite{torre} that \Pti\ 
and the average steady-state concentration of particles, $\rhoss$, 
are governed by the same critical exponent. Numerical studies have shown that this also holds for a wide variety of other models.
Assuming that this also is the case for the BAW, the following scaling
relation should hold

\begin{equation}
  \beta \ = \ \nuv \delta. \label{eq:beta}
\end{equation}

\figpinf\ shows a log-log plot of the ultimate survival probability 
\Pti\ as a function of the distance from the critical point. The 
slope of the curve yields the estimate $\beta = 0.93\pm 0.05$.

Assuming that the density of the surviving events is constant
inside a $d$-dimensional sphere expanding with constant velocity
$v$, we find that

\begin{equation}
  \ave{\Rt }^{1/2} \approx \mbox{const.}v \cdot t \label{eq:R2sup}
\end{equation}

for fixed $\Delta > 0$ and for $t \gti$. From Eq.~(\ref{eq:Rt})
we see that $\ave{\Rt }^{1/2} \propto t$ only if
$g(y) \sim y^{(1 - z/2)\nuv} \ \mbox{for} \ y \gti$, showing that

\begin{equation}
   v \propto \Delta ^{(1 - z/2)\nuv}.  \label{eq:vsup}
\end{equation}

Thus, $R_{\infty}=\lim_{t\gti}\ave{\Rt }^{1/2}/t \propto 
\Delta ^{(1-z/2)\nuv}$, so that we may obtain an estimate for 
the exponent $(1-z/2)\nuv = \nuv - \nuh$ from the asymptotic 
behavior of \Rt. In \figdsinf\ I have plotted $R_{\infty}$ as
a function of the distance from the critical point. The slope
of the line yields $\nuv - \nuh = 1.35\pm 0.10$.

Finally we may find the asymptotic behavior of the
average number of particles in the supercritical regime, as it
is simply proportional to the product of the concentration of
particles, the ultimate survival probability, and the volume of
the sphere inside which the surviving events exist. 
Eqs.~(\ref{eq:R2sup}), (\ref{eq:vsup}), and (\ref{eq:beta}) thus
gives

\begin{eqnarray}
  \nt & \approx  & \mbox{const.} \rhoss\cdot \Pti\cdot
		    (\ave{\Rt}^{1/2})^{d}     \nonumber \\
      & \sim   &  \Delta^{2\beta + (1 - z/2)\nuv d} \ t^{d}.
      \label{eq:ntsup}
\end{eqnarray}

We thus see that $n_{\infty} = \lim_{t\gti} \nt/t^{d} \propto
\Delta^{2\beta + (1 - z/2)\nuv d}$. From the scaling relation 
Eq.~(\ref{eq:zscaling}) and the conjecture $\beta/\nuh =\frac{1}{2}$
it follows that the critical exponent of $n_{\infty}$ in 1-d is
$2\beta + (1-z/2)\nuv = \nuv$. \figninf\ is a log-log plot of 
$n_{\infty}$ {\em vs} $p_{c}-p$. From the asymptotic slope of the
curve I estimate that $\nuv = 3.25 \pm 0.10$.

In the {\em subcritical} regime the process must eventually die,
and since correlations are of finite range one expects $P(t)$ to
decay {\em exponentially}. The associated scaling function must
satisfy, $\Psi(y) \propto (-y)^{\delta \nuv}\exp[-b(-y)^{\nuv}] \
\mbox{for} \ y \rightarrow  -\infty$, where $b$ is a constant.
When inserted in Eq.~(\ref{eq:sp}) this implies

\begin{equation}
P(t) \propto (-\Delta)^{\delta \nuv}\exp[-b(-\Delta)^{\nuv}t]
 \label{eq:spsub}
\end{equation}

Likewise one expects that the average number of particles decays
exponentially, $\Phi(y)\propto (-y)^{-\eta\nuv}\exp[-c(-y)^{\nuv}] \
\ \mbox{for} \ \ y \rightarrow  -\infty $, and thus

\begin{equation}
  \nt \propto (-\Delta)^{-\eta \nuv}\exp[-c(-\Delta)^{\nuv}t]
   \label{eq:ntsub}
\end{equation}

The decay constants in Eqs.~(\ref{eq:spsub}) and (\ref{eq:ntsub})
are just the inverse of the correlation time 
$\tau \propto \Delta^{-\nuv}$. We can thus find a corroborating 
estimate for $\nuv$ by studying the exponential decay of $P(t)$ and 
\nt\ in the subcritical region. In \fignpsub\ I have plotted
the estimates for $\tau$, obtained from the exponential decay of
\nt, as a function of the distance from the critical point. 
The error-bars on these data points are so large that an
improved estimate is out of the question. But clearly the
value $\nuv = \frac{13}{4}$ is fully compatible with the data.

Since the population is not growing on average, we expect the
surviving particles to spread through space via simple diffusion,
which leads to, $\Theta(y) \propto (-y)^{\nuv(1-z)} \ \mbox{for} \ \
  y \rightarrow -\infty$, which when inserted in Eq.~(\ref{eq:Rt}),
yields,

\begin{equation}
  R^{2}_{\infty} = \lim_{t \gti} \Rt/t \propto \Delta^{\nuv(1-z)}.
\end{equation}

\figr2sub\ shows a log-log plot of $R^{2}_{\infty}$ {\em vs}
$p-p_{c}$. The slope of the line is $\nuv(1-z) = \frac{13}{28}$,
which agrees with the data, though again the spread of the
data-points is quite substantial.

\subsection{Steady-state simulations}

In \figrhoss\ I have plotted \rhoss\ as a function of the distance 
from the critical point on a log-log scale, with $p_{c}=0.7215$. 
The results were obtained by averaging over typically 100 independent 
samples. The number of time steps and system sizes varied from 
$t=5000$, $L=512$ far from $p_{c}$ to $t=500000$, $L=8192$ closest to 
$p_{c}$. From the results I estimate that $\beta = 0.93(5)$ a value
consistent with the conjecture $\beta=\frac{13}{14}$.
\figchiss\ shows a plot of $\bar{\chi}$ as a function of the distance 
from the critical point on a log-lin scale, with $p_{c}=0.7215$. As
can be seen the data clearly settles down to a constant close to
the critical point thus confirming that $\gamma = 0$.

\section{Field exponent}

Finally I examine the behavior in the presence of an external 
source {\em h}, which can be added simply by allowing pairs of
nearest neighbor sites to become occupied {\em spontaneously} at 
rate {\em h}. Again, if a new particle is placed on an already
occupied site the two particles annihilate mutually. This way
of introducing the external source preserves the conservation
of particles modulo 2. The diffusion and creation processes of the BAW 
now takes place at rates $\lambda$ and unity, respectively, such 
that diffusion happens with probability $p=\lambda/(\lambda+h+1)$,
creation with probabily $q=1/(\lambda+h+1)$ and spontaneous creation
with probability $s=h/(\lambda+h+1)$. The source removes the phase
transition, just as an external magnetic field does in the Ising
model. Since $h$ is a second relevant parameter, one expects
that at the critical point, $\lambda_{c} = p_{c}/(1-p_{c}) 
\simeq 2.5907$, the order parameter decreases as a power-law
when $h \rightarrow 0$
 
\begin{equation}
  \rhoss \propto h^{1/\delta_{h}}. \label{eq:rhovsh}
\end{equation}

In \figrhovsh\ I have plotted the steady-state density of particles
as a function of $h$ at $\lambda_{c}$. The number of independent 
samples $N_{S} = 100$, the maximal number of timesteps and system 
sizes varied from $t_{M} = 1000$, $L=2048$ for large values of $h$
to $t_{M} = 250,000$, $L=8192$ for small values of $h$. The data
fit very nicely to a power-law with exponent 
$1/\delta_{h} = 0.222\pm 0.002$, which is consistent with the
conjecture $\delta_{h} = \frac{9}{2}$.

\section{Comparison with other studies \label{sec-kink}}

Grassberger {\em et al.} \cite{pgkink} studied two one-dimensional
cellular automata in which the number of kinks between ordered 
states is conserved modulo 2. Each site is in a state $S_{i} = 0,1$
and the transition rules depend only on the site itself and its
nearest neighbors. The models has the following evolution rules

\begin{center}
\begin{tabular}{lccccclcr}
$t:$ & 111 & 101 & 010 & 100 & 001 & 
$\underbrace{\mbox{\hspace{5ex}011\hspace{2ex}110\hspace{5ex}}}$ 
& 000  & \\
$t+1:$ & 0 & 0 & 1 & 1 & 1 & \raisebox{-1ex}
{$\stackrel{\textstyle 0\mbox{ with prob. }p\;\;\;\;\;\;}
{1\mbox{ with prob. }1-p}$} & 0 & model A
\end{tabular}
\end{center}

and

\begin{center}
\begin{tabular}{lccccclcr}
$t:$ & 111 & 101 & 010 & 100 & 001 & 
$\underbrace{\mbox{\hspace{5ex}011\hspace{2ex}110\hspace{5ex}}}$ 
& 000  & \\
$t+1:$ & 0 & 1 & 0 & 1 & 1 & \raisebox{-1ex}
{$\stackrel{\textstyle 1\mbox{ with prob. }p\;\;\;\;\;\;}
{0\mbox{ with prob. }1-p}$} & 0 & model B
\end{tabular}
\end{center}

When $p$ is small the system orders itself spontaneously. In model A 
there are two symmetrical absorbing states consisting of alternating
rows of 0's and 1's. In model B the time-space pattern of the 
two absorbing states form a chess-board pattern \cite{pgkink}.
When starting from a random initial configuration the system
evolves to a state with small ordered domains separated by kinks.
If $p$ is small the density of kinks decreases to zero as the system
evolves, but for $p>p_{c}$ a state with a non-zero density of kinks
is reached. The evolution rules are such that the number of kinks
is conserved modulo 2, this becomes much clearer by noting that the
evolution rules for the kinks involve the two processes 
$X \rightarrow 3X$ and $2X \rightarrow 0$. Steady-state and 
time-dependent computer simulations yielded non-DP values for various 
critical exponents. In the time-dependent simulations of model B 
Grassberger always started with a single kink so the absorbing state 
could never be reached and all trials survived indefinitely.
The simulations yielded $p_{c} = 0.5403\pm0.0013$, 
$\eta=0.272\pm0.012$, $z=1.11\pm0.02$, $\nuv=3.3\pm0.2$ and
$\beta=0.94\pm0.06$. Note that while $\eta$ is very different from
the results for the BAW, $z$ is almost the same and both $\nuv$ and 
$\beta$ are fully consistent with the estimates given above. This 
indicates that the two models belong to the same universality class, 
at least as far as the {\em static} critical behavior is concerned. 
The difference in the exponents obtained from time-dependent 
simulations are however simply due to the application of different 
initial configurations. By simulating the BAW with $n=4$ starting 
with just one particle at the origin I obtained the estimates (see
\figlsnabs) $\eta = 0.282\pm0.005$ and $z=1.140\pm 0.005$, in 
excellent agreement with Grassberger's results. It is thus clear 
that the two models belong to the same universality class.
The exponent $z$ seems to be independent of the initial configuration,
unlike $\eta$ and (off-course) $\delta$, and the exponents violate the
hyper-scaling relation Eq.~(\ref{eq:hsr}). This is reminiscent of
the situation for systems with infinitely many absorbing states
where $\delta$ and $\eta$ depend continuously on the density
of particles in the initial configuration \cite{pcpart,genscal}.
In this case a generalised scaling relation was found \cite{genscal}
to replace Eq.~(\ref{eq:hsr})

\begin{equation}
  \delta  + \beta/\nuv + \eta  = z/2, \label{eq:ghsr}
\end{equation}

where $\delta$ and $\eta$ no longer are constants and the scaling
relation, $\delta = \beta/\nuv$, is invalid. In this case I find
with $\delta = 0$, $\beta/\nuv = \frac{2}{7}$ and $z=\frac{8}{7}$
that $\eta = \frac{2}{7}$, which is consistent with the 
simulation results. 
   
Recently, Kim and Park, studied a one-dimensional monomer-dimer 
model with repulsive interactions between species of the same kind
\cite{imdfss}. When the interactions are infinitely strong 
(exclusion) Monte Carlo simulations showed a critical behavior
consistent with the BAW with 4 offspring.
In this version of the  interacting monomer-dimer model 
(IMD) a monomer $(A)$ can adsorb at an empty site only when both
nearest neighbors aren't occupied by a monomer. Likewise, a dimer
$(B_{2})$ can adsorb and dissociate onto a pair of vacancies provided 
the nearest neighbors contain no $B$-particles. There is no mutual 
exclusion between $A$ and $B$-particles. When $AB$-pairs are
formed they react immediately and the product desorbs leaving behind
two empty sites. Any state concatenated from strings of sites in the
form $A*A*$, $A*B*$ and $A*BB*$ (where $*$ indicates an empty site) is 
absorbing. The IMD has been studied using both finite-size
scaling \cite{imdfss} and time-dependent simulations \cite{imdtds}.
The results for various exponents are summarized in \tableexp. 
The critical exponents for the IMD obtained using time-dependent
simulations ($\delta$, $\eta$, $\eta '$, and $z$) agree fully
with those of the BAW with $n=4$. The exponent ratios from
finite-size scaling ($\beta/\nuh$ and $\nuv/\nuh$) are quite
similar though not in agreement to within the quoted error-bars.
Finally, the static critical exponents ($\beta$, $\nuv$ and $\nuh$)
are very different. Despite these differences I believe the IMD
is in the same universality class as the BAW with 4 offspring.
First of all time-dependent simulations are generally the most
reliable method for obtaining accurate exponent estimates, and
these exponents agree. Secondly, if one trusts the result $\eta =0$
for the IMD it follows from various scaling relations 
\footnote{For directed percolation one has \cite{pgdp}
$\gamma^{DP} = \nuv(1+\eta)$, so it follows, since 
$\gamma = \gamma^{DP}-\nuv$, that $\gamma = \nuv \eta$, the result
then follows from the arguments given in Section~3.1} (which hold
for the BAW and numerous other models) that $\beta/\nuh =\frac{1}{2}$. 
This would indicate that the IMD estimate for $\beta/\nuh$ is a
little low (or that the error-bar isn't conservative enough). One 
would therefore expect something similar to be true of the other
finite-size scaling exponent $\nuv/\nuh$, which easily could
explain the discrepancy between the IMD and BAW estimates.
Thirdly, as mentioned earlier, estimating the exponents
$\beta$, $\nuv$ and $\nuh$ is quite difficult and this difference
between the models could very well disappear with better statistics.

Why does the IMD have the same critical behavior as the BAW with
an even number of offspring and the 'kink-models' of Grassberger
{\em et al.}? The likely answer is that in all of these models
the critical dynamics is governed by the evolution 
(creation/annihilation) of domain walls between absorbing regions.
This point was stessed by Grassberger {\em et al.} \cite{pgkink} in
their analysis of their cellular automata models. Note, that the
BAW may be seen as directly modelling such domain walls. The IMD is
obviously a more complicated model and several different types
of domain walls are involved. Close to the critical point the
system consists of large regions in an absorsing configuration
separated by narrow boundary regions of 'active' sites. So even
though the different types of domain walls aren't conserved modulo 2
it still seems likely that the critical behavior is determined by the
process of getting rid of domain walls and (if possible) ending up 
with just a single absorbing domain. Though the models are quite 
different their critical behavior may very well be governed by the
same 'critical domain dynamics', which would explain why the models
exhibit the same critical behavior. These arguments are obviously
not rigorous, but still may provide some insight into why these
models apparently belong to the same universality class.

\section{Summary}

The critical exponents of the BAW with an even number of offspring
have been determined accurately and could be given exactly by
simple rational numbers listed in \tableexp. This model clearly
belongs to a non-DP universality class. I have argued that the 
cellular automata models, proposed by Grassberger {\em et al.}
\cite{pgkink}, for the dynamics of kinks between ordered (absorbing) 
states and a recently proposed interacting monomer-dimer model
\cite{imdfss} also belong to this universality class. The novel
non-DP critical behavior is apparently associated with the 
process of eliminating domain walls (kinks) in order to end
up with a single absorbing domain.

{\Large \bf Acknowledments}

I am grateful to M. H. Kim and H. Park for communicating their
results prior to publication. I would like to thank R. Dickman and
H. Park for interesting and enlightning discussions on the BAW and
IMD models.

\newpage

{\Large \bf Tables}

{\bf \tableexp:} Simulation results and conjectured values for the 
critical exponents for the one-dimensional BAW with 4 offspring
from this work and simulation results for the interacting 
monomer-dimer model from Ref.'s~\cite{imdfss,imdtds}.
$\eta '$ is the exponent governing the growth of the average
number of particles when the absorbing can't be reached. 
For comparison is listed the corresponding exponent estimates
for models in the DP universality class, taken from 
Ref.'s~\cite{tdpjsp,baxterdp} ($\beta$), 
\cite{essamdp} ($\gamma$, $\nuv$, and $\nuh$) and 
\cite{fielddp} ($\delta_{h}$). In the estimate for $\gamma$ I
used $\gamma = \gamma^{DP}-\nuv$. The estimates for the
remaining exponents were obtained using scaling relations.
The figures in parenthesis are the quoted uncertainties in the
last digits.
\begin{center}
\begin{tabular}{ccccc}
\hline \hline
 Exponent & BAW Simulations & Conjecture & IMD simulations & DP class \\
\hline
$\beta$ & 0.92(3) & $\frac{13}{14}$ & 0.78(3) & 0.2767(4)\\
$\gamma$ & 0.00(5) & 0 & N.A. & 0.5438(13) \\
$\nuv$  & 3.25(10) & $\frac{13}{4}$ & 2.88(5) & 1.7334(10)\\
$\nuh$  & 1.84(6) & $\frac{13}{7}$ & 1.73(3) & 1.0972(6) \\
$\beta/\nuh$ & 0.500(5)& $\frac{1}{2}$ & 0.45(2) & 0.2522(6) \\
$\nuv/\nuh$ & 1.750(5) & $\frac{7}{4}$ & 1.664(3) & 1.5798(18) \\
$\delta$ & 0.285(2)& $\frac{2}{7}$& 0.30(2) &0.1596(4) \\
$\eta$ & 0.000(1) & 0 &  0.00(2) & 0.3137(10) \\
$\eta'$ & 0.282(4) & $\frac{2}{7}$ & 0.28(3)  & ------\\
$z$  & 1.141(2) & $\frac{8}{7}$ & 1.25(15)  & 1.2660(14) \\
$1/\delta_{h}$ & 0.222(2) & $\frac{2}{9}$ &  N.A. &0.111(3) \\
\hline \hline
\end{tabular}
\end{center}

\newpage

{\Large \bf Figure Captions}
  
{\bf \figlsfar:}  Local slopes $-\delta (t)$ (upper panel), 
$\eta (t)$ (middle panel), and $z(t)$ (lower panel), as defined in 
Eq.~\protect{\ref{eq-ls}} with $m=5$. Each panel contains five curves 
with, from top to bottom, $p = 0.715$, 718, 721, 724, and 727.

{\bf \figlsclose:} Same as in \figlsfar\ but for $p = 0.7210$, 7215, 
and 7220.

{\bf \figrhovsl:} $\rho_{s}(p_{c},L)$ {\em vs} $L$ at 
$p_{c} = 0.7215$. The slope of the line is $\beta/\nuh=\frac{1}{2}$.
                     
{\bf \figchivsl:} Log-lin plot of the susceptibility 
$\chi_{s}(p_{c},L)$ {\em vs} $L$ at the critical point 
$p_{c} = 0.7215$. 
                  
{\bf \figtauvsl:} $\tau(p_{c},L)$ vs $L$ with 
$p_{c} = 0.7215$. The slope of the line is $y=\nuv/\nuh=\frac{7}{4}$.
           
{\bf \figrhoinit:} The short-time decay of the density of particles 
$\rho(p_{c},L,t)$ as a function of $t$ at the critical point with 
$L=2^{15}$. The slope of the line is $\theta=\beta/\nuv=\frac{2}{7}$.

{\bf \figsupscale:} Plots of $\Delta^{-\delta \nuv}P(t)$ 
(upper panel), $\Delta^{\eta \nuv}\nt$ (middle panel), and
$\Delta^{z\nuv}\Rt$ (lower panel) {\em vs} $\Delta t^{1/\nuv}$
for various values of $p$ in the supercritical region. 
The critical parameters are $p_{c}=0.7215$,
$\nuv = \frac{13}{4}$, $\delta = \frac{2}{7}$, $\eta = 0$, and 
$z = \frac{8}{7}$. In each panel is an inset showing the unscaled 
data, i.e., $P(t)$, \nt, and \Rt\ vs $1/t$.

{\bf \figpinf:} The ultimate survival probability $\Pti$ {\em vs} 
$p_{c}-p$. The slope of the line is 
$\beta = \frac{13}{14}$.
                 
{\bf \figdsinf:}  $R_{\infty} = \lim_{t \gti} \ave{\Rt}^{1/2}/t$ 
{\em vs} $p_{c}-p$. The slope of 
the line is $\nuv - \nuh = \frac{39}{28}$.

{\bf \figninf:} The ultimate number of particles $n_{\infty}  = 
\lim_{t  \gti} \nt/t$ {\em vs} $p_{c}-p$. 
The slope of the line is $\nuv = \frac{13}{4}$.

{\bf \fignpsub:} The correlation time $\tau$ governing the
decay of the number of particles {\em vs} $p-p_{c}$. The slope 
of the line is $\nuv = \frac{13}{4}$.

{\bf \figr2sub:} Log-log plot of $R^{2}_{\infty}$ {\em vs}
$p-p_{c}$. The slope of the line is $\nuv(1-z) = \frac{13}{28}$.

{\bf \figrhoss:} The steady-state density of particles $\rho$
{\em vs} $p_{c}-p$. The slope of the straight line is 
$\beta=\frac{13}{14}$.

{\bf \figchiss:} The steady-state fluctuations in the number 
of particles $\chi$ {\em vs} the distance from the critical point. 

{\bf \figrhovsh:} The steady-state density of particles as a function
of the external source $h$ at the critical point $p_{c} = 0.7215$.
The slope of the line is $1/\delta_{h} = \frac{2}{9}$.

{\bf \figlsnabs:} Local slopes $\eta (t)$ and $z (t)$ at
$p = 0.7215$ starting with a single particle.

\newpage
\begin{thebibliography}{99}
\bibitem{dpbook} Several articles about directed percolation may 
be found in {\em Percolation Structures and Processes}, edited by
G. Deutscher, R. Zallen, and J. Adler, Annals of the Israel Physical
Society  Vol. 5 (Hilger, Bristol, 1983).

\bibitem{cardy} J. L. Cardy and R. L. Sugar, J. Phys. A \bf13\rm, 
L423 (1980)

\bibitem{janssen2} H. K. Janssen, Z. Phys. B {\bf 58}, 311 (1985)

\bibitem{pgdp} P. Grassberger, J. Phys. A {\bf 22}, 3673 (1989)

\bibitem{harris} T. E. Harris, Ann. Prob. {\bf 2}, 969 (1974)

\bibitem{liggett} T. M. Liggett, {\em Interacting Particle
Systems} (Springer-Verlag, New York, 1985).

\bibitem{durrett} R. Durrett, {\em Lecture Notes on Par\-ticle
Sy\-stems and Per\-co\-la\-tion} (Wadsworth Pub. Co.,
Pacific Grove, CA, 1988)

\bibitem{schlogl} F. Schl\"{o}gl, Z. Phys. B {\bf 253}, 147 (1972)

\bibitem{torre} P. Grassberger and A. de la Torre, Ann. Phys. (NY)
 {\bf 122}, 373 (1979)

\bibitem{janssen1} H. K. Janssen, Z. Phys. B {\bf 42}, 151 (1981)

\bibitem{pg2} P. Grassberger, Z. Phys. B {\bf 47}, 365 (1982)

\bibitem{brower} R. C. Brower, M. A. Furman, and M. Moshe,
Phys. Lett. B {\bf 76}, 213 (1978)

\bibitem{bbc} R. Bidaux, N. Boccara, and H. Chat\'{e}, 
Phys. Rev. A {\bf 39}, 3094 (1989). \newline
I. Jensen, Phys. Rev. A {\bf 43}, 3187 (1991).

\bibitem{aukrust} T. Aukrust, D. A. Browne, and I. Webman,
Phys. Rev. A  {\bf 41}, 5294 (1990)

\bibitem{tripann} R. Dickman, Phys. Rev. A {\bf 42}, 6985 (1990).

\bibitem{tripcrea} R. Dickman and Tania Tom\'{e},
Phys. Rev. A {\bf 44}, 4833 (1991).

\bibitem{rondif} R. Dickman, Phys. Rev. B {\bf 40}, 7005 (1989).

\bibitem{zgb} R. M. Ziff, E. Gulari, and Y. Barshad,
Phys. Rev. Lett. {\bf 56}, 2553 (1986). \newline
G. Grinstein, Z.-W. Lai, and D. A. Browne,
Phys. Rev. A {\bf 40}, 4820 (1989). \newline
I. Jensen, H. C. Fogedby, and R. Dickman,
Phys. Rev. A {\bf 41}, 3411 (1990).

\bibitem{pcpart} I. Jensen,  Phys. Rev. Lett. {\bf 70}, 1465 (1993). 
\newline I. Jensen and R. Dickman, Phys. Rev. E. {\bf 48}, 1710 (1993).

\bibitem{bawprl} H. Takayasu and A. Yu. Tretyakov,
Phys. Rev. Lett. {\bf 68}, 3060 (1992).

\bibitem{bawbramson} M. Bramson and L. Gray,
Z. Warsch. verw. Gebiete {\bf 68}, 447 (1985).

\bibitem{tdpjsp} I. Jensen and R. Dickman, J. Stat. Phys.
{\bf 71}, 89 (1993).
		       
\bibitem{iwanbaw} I. Jensen, Phys. Rev. E {\bf 47}, R1 (1993).

\bibitem{inuicam} N. Inui, Phys. Lett. A {\bf 184}, 79 (1993).

\bibitem{cam} M. Suzuki, J. Phys. Soc. Jpn. {\bf 55}, 4205 (1986).
\\
M. Suzuki, M. Katori and X. Hu, J. Phys. Soc. Jpn. {\bf 56}, 
3092 (1987). \\
M. Katori and M. Suzuki, J. Phys. Soc. Jpn. {\bf 56}, 3113 (1987).

\bibitem{bawsudbury} A. Sudbury, Ann. Probab. {\bf 18}, 581 (1990).

\bibitem{bawtakayasu} H. Takayasu and N. Inui, J. Phys. A {\bf25},
L585 (1992).

\bibitem{pgkink} P. Grassberger, F. Krause, and T. von der Twer,
J. Phys. A {\bf 17}, L105 (1984). \\
P. Grassberger, J. Phys. A {\bf 22}, L1103 (1989).

\bibitem{bawann} I. Jensen, J. Phys. A {\bf 26}, 3921 (1993).

\bibitem{cp3d} I. Jensen, Phys. Rev. A. {\bf 46}, 7393 (1992).

\bibitem{fss} M. E. Fisher, in {\em Procedings of the Enrico
Fermi International School of Physics, Vol. 51}, edited by M. S. Green
(Academic Press, Varenna, Italy, 1971). \\
M. N. Barber, in {\em Phase Transitions and Critical
Phenomena, Vol. 8}, edited by C. Domb and J. L. Lebowitz, 
(Academic Press, New York, 1983). \\
V. Privman, Ed., {\em Finite Size Scaling and Numerical Simulations
of Statistical Systems}, (World Scientific, Singapore, 1990).

\bibitem{genscal} J. F. F. Mendes, R. Dickman, M. Henkel and
M. C. Marques, ``Generalized scaling for models with multiple 
absorbing states", preprint 1993, cond-mat/9312028.

\bibitem{imdfss} M. H. Kim and H. Park, ``Critical behavior of
an interacting monomer-dimer model", preprint 1993, cond-mat/9312048.

\bibitem{imdtds} H. Park, private communication.

\bibitem{baxterdp} R. J. Baxter and A. J. Guttmann,
J. Phys. A {\bf 21}, 3193 (1988).

\bibitem{essamdp} J. W. Essam, K. De'Bell, J. Adler and
F. M. Bhatti, Phys. Rev. B {\bf33}, 1982 (1986). \\
J. W. Essam, A. J. Guttmann and K. De'Bell,
J. Phys. A {\bf 21}, 3815 (1988).


\bibitem{fielddp} J. Adler and J. A. M. S. Duarte, 
Phys. Rev. B {\bf 35}, 7046 (1987).


\end{thebibliography}
\end{document}